\documentclass[10pt,conference]{IEEEtran}
\usepackage{subfigure}
\usepackage{amssymb}
\usepackage{amsthm}
\usepackage{amsmath}
\usepackage{hyperref}
\usepackage{graphicx}
\usepackage{colortbl,dcolumn}
\usepackage{booktabs}
\usepackage{xcolor}
\usepackage{cite}
\usepackage{threeparttable}

\usepackage{algorithm}
\usepackage{algorithmic}

\newtheorem{definition}{Definition}

\newtheorem{theorem}{Theorem}
\newtheorem{proposition}{Proposition}

\newtheorem{remark}{Remark}
\newtheorem{problem}{Problem}

\newcommand{\defref}[1]{Definition~\ref{#1}}

\newcommand{\thmref}[1]{Theorem~\ref{#1}}
\newcommand{\propref}[1]{Proposition~\ref{#1}}

\newcommand{\figref}[1]{Fig.~\ref{#1}}
\newcommand{\tabref}[1]{Table~\ref{#1}}
\newcommand{\secref}[1]{Section~\ref{#1}}

\newcommand{\probref}[1]{Problem~\ref{#1}}
\newcommand{\rekref}[1]{Remark~\ref{#1}}
\newcommand{\algref}[1]{Algorithm~\ref{#1}}

\usepackage{mathtools} 

\usepackage{enumerate}

\ifCLASSOPTIONcompsoc
\else
\fi
\ifCLASSINFOpdf
\else
\fi

\graphicspath{{pdfs/},{images/}}

\begin{document}
%
\title{Rate-Achieving Policy in Finite-Horizon Throughput Region for Multi-User Interference Channels}
%
%
%
%

\author{\IEEEauthorblockN{Yirui Cong, Xiangyun Zhou, and Rodney A. Kennedy}\\
\IEEEauthorblockA{Research School of Engineering, The Australian National University, Canberra, ACT 0200, Australia\\
Emails: \{yirui.cong, xiangyun.zhou, Rodney.Kennedy\}@anu.edu.au}
}

\IEEEtitleabstractindextext{%
\begin{abstract}

This paper studies a wireless network consisting of multiple transmitter-receiver pairs sharing the same spectrum where interference is regarded as noise. Previously, the throughput region of such a network was characterized for either one time slot or an infinite time horizon. This work aims to close the gap by investigating the throughput region for transmissions over a finite time horizon. We derive an efficient algorithm to examine the achievability of any given rate in the finite-horizon throughput region and provide the rate-achieving policy. The computational efficiency of our algorithm comes from the use of A* search with a carefully chosen heuristic function and a tree pruning strategy. We also show that the celebrated max-weight algorithm which finds all achievable rates in the infinite-horizon throughput region fails to work for the finite-horizon throughput region.

\end{abstract}

}

\maketitle

\IEEEdisplaynontitleabstractindextext

%
\IEEEpeerreviewmaketitle

\section{Introduction}

\subsection{Motivation and Related Work}

Analyzing the throughput region under any given modulation and coding strategy is an important issue for studying the network capacity from a network-layer perspective~\cite{NeelyM2005JSAC}. Such studies commonly assume that the interference in the network is treated as noise, hence the capacity of each link is determined by signal-to-interference-plus-noise ratio (SINR).
In this work, we take the same network-layer approach and study the throughput region of a wireless network having multiple transmitter-receiver pairs.
%
%
The key difference between point-to-point systems and multi-user networks is the consideration of multiple time slots. For point-to-point systems, knowing the achievable rate and the rate-achieving transmission policy in one time slot is sufficient to derive the rate-achievable results for any number of time slots. However, this is not the case for multi-user networks, where the throughput region over multiple time slots is different from that in a single time slot. In fact, the multi-slot throughput region is generally larger than the single-slot throughput region~\cite{GeorgiadisL2006BOOK,VaeziM2016ICC}.

A number of studies investigated the achievable rates in multi-user wireless networks over an infinite number of time slots. The seminal work for infinite-horizon throughput region\footnote{In this paper, the term `infinite horizon' refers to an infinite number of time slots and the term `finite horizon' refers to a finite number of time slots.} was introduced in~\cite{TassiulasL1992,TassiulasL1991Thesis} and further generalized in~\cite{GeorgiadisL2006BOOK,NeelyM2005JSAC,LinX2006,NeelyM2010,LeL2012,XueD2013VT,XueD2015}. These studies revealed the relationship between the exogenous data rate, which is the rate at which data arrives in the data queue of each transmitter, and the infinite-horizon throughput region formed by all the achievable rates over an infinite number of time slots. If a given exogenous rate is in the infinite-horizon throughput region, there exists a rate-achieving transmission policy to result in a stable data queue condition. It is also shown that the infinite-horizon throughput region is convex~\cite{TassiulasL1992,TassiulasL1991Thesis,GeorgiadisL2006BOOK,NeelyM2005JSAC}.

Despite the theoretical importance of the infinite-horizon throughput region result, it does not provide sufficient insights into the throughput region or rate-achieving policy over a finite horizon, i.e., a finite number of time slots. In wireless networks, the network traffic, channel condition and even network topology change with time~\cite{GeorgiadisL2006BOOK}. Transmission should always be designed for a finite time duration, i.e., a relatively small number of time slots, such that the network and channel information used in the design is not outdated when the actual transmission happens. In addition, achieving real-time quality of service (QoS) also requires design over a finite horizon instead of an infinite horizon. To the best of our knowledge, the finite-horizon throughput region of a multi-user wireless network has not yet been investigated.

\subsection{Our Contributions}

In this work, we investigate the finite-horizon throughput region of a wireless network consisting of multiple transmitter-receiver pairs. Our approach is not to completely characterize the finite-horizon throughput region because unlike the infinite counterpart, it is non-convex and the complexity of finding all achievable rates increases exponentially with the number of time slots. Instead, we provide a method to determine (i) whether an arbitrarily given rate is achievable, and (ii) if so, what the rate-achieving transmission policy is. We formulate the problem of finding the rate-achieving policy in terms of the transmission-time-minimization problem and provide an efficient solution based on an interference-free based heuristic function. We prove this heuristic function is admissible so that the celebrated A* search algorithm can be implemented~\cite{RussellS2009BOOK}, which largely improves the computational efficiency.

We also highlight a fundamental difference between finite-horizon throughput region and the previously studied infinite-horizon throughput region. Specifically, we show that the well-known max-weight algorithm~\cite{TassiulasL1992} which can achieve all rates in the interior of the infinite-horizon throughput region fails to find the achievable rate in the interior of the finite-horizon throughput region. This suggests that the existing methods dealing with the rate-achieving policies for infinite horizon cannot be directly applied to study the case of finite horizon.

%

\subsection{Notation}

Throughout this paper, for a vector $\mathbf{a} = [a^{(1)},\ldots,a^{(N)}]^{\mathrm{tr}}$ (where $\mathrm{tr}$ denotes the transpose operator), $(\mathbf{a})^+$ denotes $\max\{a^{(n)},0\}$ for all $n\in\{1,\ldots,N\} =: \mathcal{N}$.
The cardinality of a set $\mathcal{A}$ is $|\mathcal{A}|$.
For $\mathbf{x}_1 = [x_1^{(1)},\ldots.,x_1^{(N)}]^{\mathrm{tr}}$ and $\mathbf{x}_2 = [x_2^{(1)},\ldots,x_2^{(N)}]^{\mathrm{tr}}$, $\mathbf{x}_1\succeq$($\succ,\preceq,\prec$) $\mathbf{x}_2$ represents $x_1^{(n)}\geq$($>,\leq,<$) $x_2^{(n)}$ for all $n\in\mathcal{N}$.
$\overline{\mathbb{R}}_{+}^{N}$ ($\mathbb{R}_+^N$) means $\left\{\mathbf{x} \in \mathbb{R}^N\colon \mathbf{x} \succeq \, (\succ)~0\right\}$.
And $\mathbf{0}$ stands for the zero vector.

\section{System Model and Problem Description}\label{sec:System Model and Problem Description}

\subsection{System Model}\label{sec:System Model}

\begin{figure}[h]
\centering
\includegraphics [width=0.7\columnwidth]{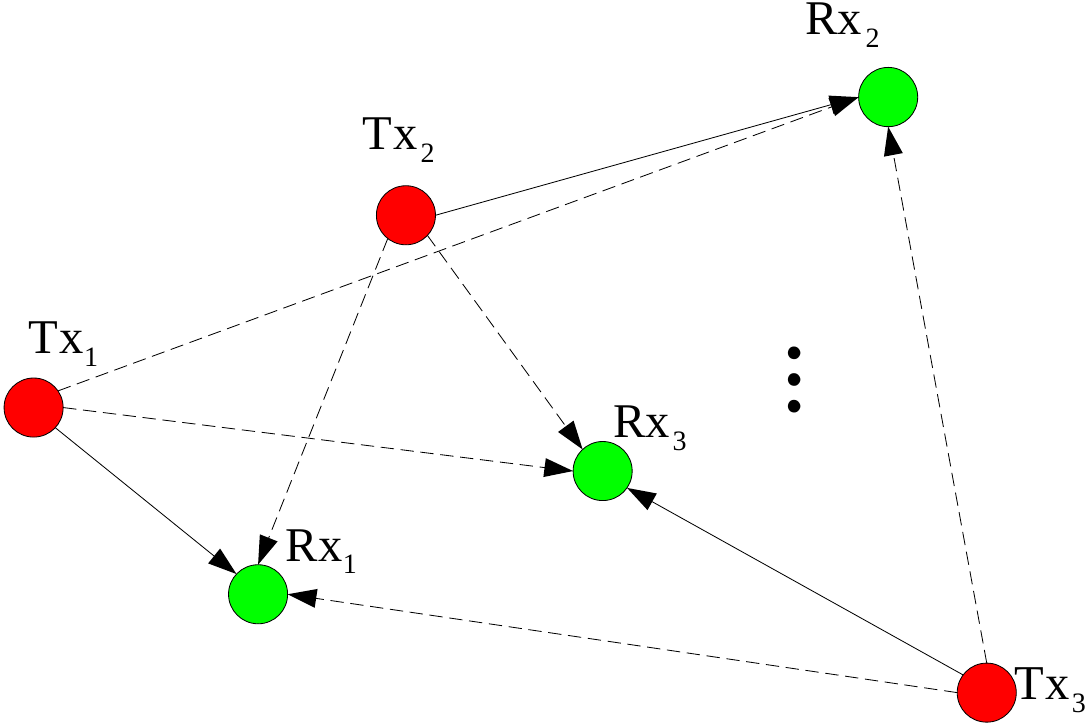}
\caption{Transmitter-receiver pairs in a wireless network.
The solid and dashed lines indicate the desired data signals and the interference signals, respectively.}
\label{fig:Transmitter and receiver pairs in wireless networks}
\end{figure}

Assume there are $N$ transmitter-receiver pairs sharing the same bandwidth in a wireless network, as shown in \figref{fig:Transmitter and receiver pairs in wireless networks}.
Specifically, $\mathrm{Tx}_n$ and $\mathrm{Rx}_n$ denote the transmitter and receiver of the $n$\textsuperscript{th} communication pair.
The power gain of the channel between $\mathrm{Tx}_n$ and $\mathrm{Rx}_m$ is denoted by $h_{nm}$. All power gains remain constant for a given finite time horizon.
The time is slotted, and each time slot is the period of transmitting and receiving a codeword.
We consider a finite time horizon of $T$ time slots, which is no longer than the channel coherent time, and the duration of each slot is $\tau$.

In each time slot, every transmitter-receiver pair chooses to transmit or not.
That is, for time slot $t\in\{1,\ldots,T\}=:\mathcal{T}$, the transmitter $\mathrm{Tx}_n$ ($n\in\{1,\ldots,N\}=:\mathcal{N}$) can choose its transmit power $s_t^{(n)}$ from the transmit-power set $\mathcal{S}^{(n)}$, in which $0$ is included for representing no transmission.
%
%
Since the number of available power options in a practical communication system is usually finite, we model $\mathcal{S}^{(n)}$ as a finite set.
Furthermore, we label $\mathbf{s}_t = \big[s_t^{(1)},\ldots,s_t^{(N)}\big]^{\mathrm{tr}}$, and $\mathcal{S} := \mathcal{S}^{(1)}\times\dotsb\times\mathcal{S}^{(N)}$.
Hence, $\mathbf{s}_t \in \mathcal{S}$ and we call $\mathcal{S}$ the transmit-power-vector set.

For time slot $t$, the SINR for each transmitter-receiver pair is determined by
\begin{align}\label{eqn:SINR}
\gamma_n(\mathbf{s}_t) = \frac {h_{nn} s_t^{(n)}} {W_n + \sum_{m\neq n} h_{mn} s_t^{(m)}}, \quad n,m\in\mathcal{N},
\end{align}
where $W_n$ is the power of additive white Gaussian noise for $\mathrm{Rx}_n$ during transmission.
The capacity of $N$ transmitter-receiver pairs by applying power vector $\mathbf{s}_t$ is\footnote{As discussed in the introduction, we do not consider information-theoretic capacity.
The capacity definition in~\eqref{eqn:Capcities for a Given Power} is given in~\cite{NeelyM2005JSAC,GeorgiadisL2006BOOK} and implicitly assumes that the interference is treated as noise.}
\begin{align}\label{eqn:Capcities for a Given Power}
\mathbf{C}(\mathbf{s}_t) \!=\! \left[\log_2 \left(1 \!+\! \frac {\gamma_1(\mathbf{s}_t)} {\Gamma_1}\right),\dotsc,\log_2 \left(1 \!+\! \frac {\gamma_N(\mathbf{s}_t)} {\Gamma_N}\right)\right]^{\mathrm{tr}}
\end{align}
where $\Gamma_n \geq 1$ ($n \in \mathcal{N}$) represents generally any gap to capacity \cite{TanC2013} due to practical finite blocklength coding and practical modulation schemes.
We absorb $1/\Gamma_n$ into $h_{nn}$ and thus~\eqref{eqn:Capcities for a Given Power} can be rewritten as
\begin{align}\label{eqn:Equivalent Capcities for a Given Power}
\mathbf{C}(\mathbf{s}_t) = \left[\log_2 (1 + \gamma_1(\mathbf{s}_t)),\dotsc,\log_2 (1 + \gamma_N(\mathbf{s}_t))\right]^{\mathrm{tr}}.
\end{align}
We say a rate ${\boldsymbol \mu}_t \in \overline{\mathbb{R}}_+^N$ (in time slot $t$) is achievable when ${\boldsymbol \mu}_t \preceq C(\mathbf{s}_t)$.
For time slot $t$, all achievable rates form a one-slot throughput region
\begin{align}\label{eqn:One-Slot Throughput Region}
\Lambda_{[1],t} = \bigcup_{\mathbf{s}_t \in \mathcal{S}}\left\{{\boldsymbol \mu}_t\colon 0 \preceq {\boldsymbol \mu}_t \preceq \mathbf{C}(\mathbf{s}_t)\right\}.
\end{align}
Note that $\Lambda_{[1],t}$ are the same for all $t$, and thus, for simplicity, we label $\Lambda_{[1],1} = \dotsb = \Lambda_{[1],T} = \Lambda_{[1]}$.
%

Similar to the one-slot throughput region, the finite-horizon throughput region for $T$ time slots is defined as follows.

\begin{definition}[Finite-Horizon Throughput Region]\label{def:Finite-Slots Throughput Region}
The $T$-slot throughput region $\Lambda_{[T]}$ is the set of average rates that can be achieved in $T$ time slots, i.e.,
\begin{align}\label{eqn:Finite-Horizon Throughput Region}
\Lambda_{[T]} = \left\{{\boldsymbol \mu}_{[T]}\colon {\boldsymbol \mu}_{[T]} = \frac{1}{T}\sum_{t=1}^{T} {\boldsymbol \mu}_t,\quad{\boldsymbol \mu}_t \in \Lambda_{[1]}\right\}.
\end{align}
\end{definition}


We also define the weak Pareto frontier and Pareto frontier, which are very helpful in the later parts of the paper.

\begin{definition}[Weak Pareto Frontier and Pareto Frontier]\label{def:Weak Pareto Frontier and Pareto Frontier}
For a set $\mathcal{A}$, the weak Pareto frontier is
\begin{align}\label{eqn:Weak Pareto Frontier}
\mathcal{B} = \left\{b\in\mathcal{A}\colon\left\{a\in\mathcal{A}:a \succ b \right\} = \emptyset\right\},
\end{align}
and the Pareto Frontier is
\begin{align}\label{eqn:Pareto Frontier}
\overline{\mathcal{B}} = \left\{b\in\mathcal{A}\colon\left\{a\in\mathcal{A}\colon a \succeq b \right\} = \{b\}\right\}.
\end{align}
It should be noted that $\overline{\mathcal{B}} \subseteq \mathcal{B}$.
\end{definition}

\begin{figure}[h]
\centering
\includegraphics [width=0.7\columnwidth]{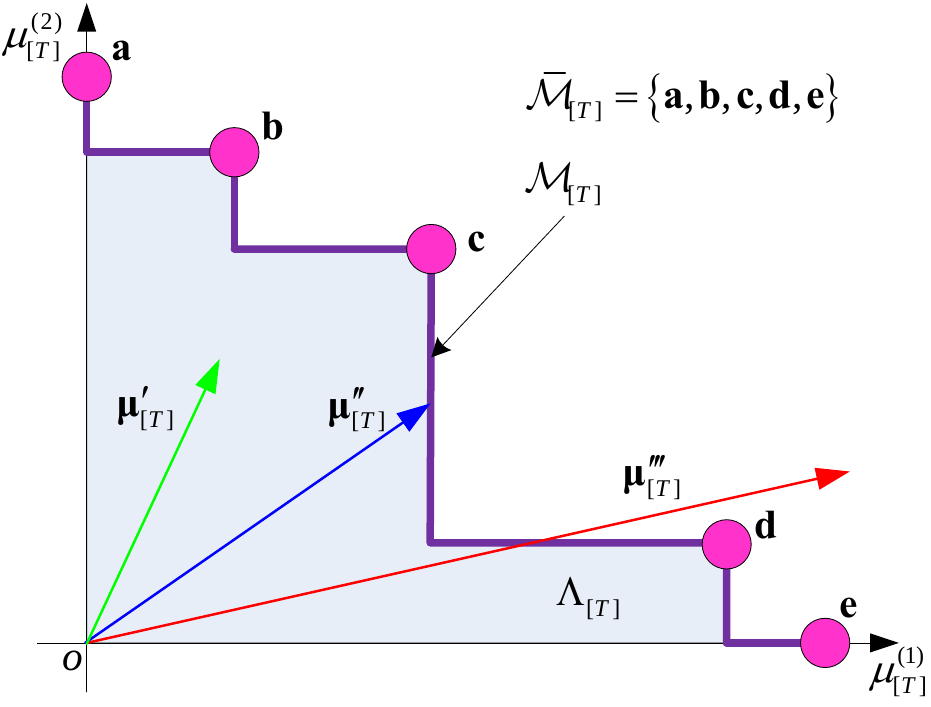}
\caption{The Pareto frontier and weak Pareto frontier of throughput region.
For two transmitter-receiver pairs, we have ${\boldsymbol \mu}_{[T]} = \big[\mu^{(1)}_{[T]},\mu^{(2)}_{[T]}\big]^{\mathrm{tr}}$ and the $T$-slot throughput region is in a two-dimensional space.
The purple (thick) line segments stand for the weak Pareto frontier $\mathcal{M}_{[T]}$ for $\Lambda_{[T]}$, and the points $\mathbf{a}, \mathbf{b}, \mathbf{c}, \mathbf{d}, \mathbf{e}$ collectively make the Pareto frontier $\overline{\mathcal{M}}_{[T]}$ of $\Lambda_{[T]}$.
${\boldsymbol \mu}'_{[T]}$ is in $\Lambda_{[T]}\setminus\mathcal{M}_{[T]}$, and ${\boldsymbol \mu}''_{[T]}$ is in $\mathcal{M}_{[T]}$, but ${\boldsymbol \mu}'''_{[T]}$ is not in $\Lambda_{[T]}$.
}
\label{fig:Illustrations on Pareto Frontier and Weak Pareto Frontier of Throughput Region}
\end{figure}

With \defref{def:Weak Pareto Frontier and Pareto Frontier}, we define the weak Pareto frontier and Pareto frontier of $\Lambda_{[1]}$ as $\mathcal{M}_{[1]}$ and $\overline{\mathcal{M}}_{[1]}$, respectively.
Similarly, $\mathcal{M}_{[T]}$ and $\overline{\mathcal{M}}_{[T]}$ stand for the weak Pareto frontier and Pareto frontier of $\Lambda_{[T]}$.
%
%
\figref{fig:Illustrations on Pareto Frontier and Weak Pareto Frontier of Throughput Region} gives a pictorial illustration on $\Lambda_{[T]}$, $\mathcal{M}_{[T]}$ and $\overline{\mathcal{M}}_{[T]}$. It is clear that the finite-horizon throughput region is generally non-convex.\footnote{Note that the throughput region is different from that using the time-sharing method in~\cite{MotahariA2009}, where the length of the ``time slot'' can be arbitrarily selected which is impractical.} This is in contrast to the infinite-horizon throughput region which is convex.

\subsection{Problem Description}\label{sec:Problem Description}

This work focuses on how to achieve any given rate in $\Lambda_{[T]}$.
To achieve a rate, say ${\boldsymbol \mu}_{[T]}$, we need to determine the \textit{transmission rate} and \textit{power} in every time slot, which gives the rate-achieving policy. The exact definition of the rate-achieving policy is given as follows.



\begin{definition}[Rate-Achieving Policy]\label{def:Rate-Achieving Policy}
For a given transmit-power-vector set $\mathcal{S}$ and a finite horizon of $T$ time slots, $\forall {\boldsymbol \mu}_{[T]} \in \Lambda_{[T]}$, the rate-achieving policy for ${\boldsymbol \mu}_{[T]}$ is a sequence of rate-power pairs
\begin{align}\label{eqn:Rate-Achieving Policy}
\mathcal{P}_{T} = \left({\boldsymbol \mu}_t, \mathbf{s}_t\right)_{t=1}^{T},\,\mathbf{s}_t \in \mathcal{S},
\end{align}
with the capacity constraint\footnote{For each transmitter-receiver pair in every time slot, the transmission rate should not exceed the corresponding capacity.} ${\boldsymbol \mu}_t \preceq \mathbf{C} \left(\mathbf{s}_t\right)$ such that ${\boldsymbol \mu}_{[T]}$ can be achieved, i.e.,
\begin{align}\label{eqn:Rate Achievement Equation}
{\boldsymbol \mu}_{[T]} = \frac{1} {T} \sum_{t=1}^{T} {\boldsymbol \mu}_t.
\end{align}
\end{definition}


The main task of this paper is to develop a computationally efficient way to find the rate-achieving policy for any achievable rate. Our result will also tell whether a given rate is achievable or not.

\section{Main Results}\label{sec:Main Results}

To find the rate-achieving policy, we define the following equivalent transmission-time-minimization problem (see \probref{prob:Transmission-Time-Minimization Problem}).
The main idea for establishing this equivalent problem is: achieving a given average rate ${\boldsymbol \mu}_{[T]}$ over $T$ time slots is the same as transmitting $\tau T{\boldsymbol \mu}_{[T]}$ amount of data within $T$ time slots, where $\tau$ is the length of each time slot.

\begin{problem}\label{prob:Transmission-Time-Minimization Problem}
For a given transmit-power-vector set $\mathcal{S}$ and a finite horizon of $T$ time slots, the equivalent transmission-time-minimization problem is
\begin{align}\label{eqn:Transmission-Time-Minimization Problem}
\!\!\!\!\!\!\!\!\begin{array}{l l}
\underset {\left(\mathbf{s}_t\right)_{t=1}^{p},\,\mathbf{s}_t\in\mathcal{S}} {\mathrm{minimize}} & p\\
\mathrm{subject~to}~&\mathbf{Q}_{t} = \left(\mathbf{Q}_{t-1} \!-\! \tau \mathbf{C}\left(\mathbf{s}_t\right)\right)^+,\,t\!\in\!\{1,\ldots,p\},\\
&\mathbf{Q}_{0} = \tau T {\boldsymbol \mu}_{[T]},\\
&\mathbf{Q}_{p} = \mathbf{0},
\end{array}
\end{align}
where $p$ denotes the number of time slots for completing the transmission and is a variable dependent on $\left(\mathbf{s}_t\right)_{t=1}^{p}$.
Additionally, $\mathbf{Q}_t = [Q_t^{(1)},\ldots,Q_t^{(N)}]^{\mathrm{tr}}$ in~\eqref{eqn:Transmission-Time-Minimization Problem}, and each $Q_t^{(n)} \in \overline{\mathbb{R}}_+$ is the length of an equivalent virtual data queue in transmitter $\mathrm{Tx}_n$ after $\mathbf{s}_t$ is applied in time slot $t$ ($t \in \{1,\ldots,p\}$). The vector
$\mathbf{Q}_0$ contains the lengths of the initial data queues before applying $\mathbf{s}_1$. The vector
${\boldsymbol \mu}_{[T]} = [\mu_{[T]}^{(1)},\ldots,\mu_{[T]}^{(N)}]^{\mathrm{tr}}$ is the given data rate to be achieved.
A solution of optimal design parameters (not unique for $T > 1$) is denoted as $(\mathbf{s}_t^*)_{t=1}^{p^*}$.
We label the optimal objective as $p^*$, which stands for the minimum number of time slots to clear the data queue.
The corresponding vector of data-queue sequence under the optimal solution is denoted by $(\mathbf{Q}_t^*)_{t=1}^{p^*}$.
\end{problem}

In the rest of this section, we will give detailed discussions on how to derive the rate-achieving policy (see \secref{sec:Deriving Rate-Achieving Policy}) based on the solution of \probref{prob:Transmission-Time-Minimization Problem}. A computationally efficient algorithm for solving \probref{prob:Transmission-Time-Minimization Problem} will be presented in \secref{sec:Solution Method for Transmission-Time-Minimization Problem}.

\subsection{Deriving the Rate-Achieving Policy}\label{sec:Deriving Rate-Achieving Policy}

In this subsection, we derive the rate-achieving policy for any given achievable rate.
It should be noted that our method is complete, i.e., for any given achievable rate, the corresponding rate-achieving policy can be obtained.
In contrast, the classical max-weight algorithm~\cite{TassiulasL1992} is not complete, which is discussed at the end of this subsection.

First, we present the rate-achieving policy for all rates in the $T$-slot throughput region as follows.

\begin{theorem}[Rate-Achieving Policy]\label{thm:Rate-Achieving Policy for All Achievable Rates}
Given a transmit-power-vector set $\mathcal{S}$ and a finite horizon of $T$ time slots, then:
\begin{enumerate}[i)]
\item   If ${\boldsymbol \mu}_{[T]} \in \Lambda_{[T]}$, then $p^* \leq T$, and the rate-achieving policy is $\mathcal{P}_T = ({\boldsymbol \mu}_t, \mathbf{s}_t)_{t=1}^T$ with
\begin{align}\label{eqn:Rate-Achieving Policy for All Achievable Rates}
\left({\boldsymbol \mu}_t, \mathbf{s}_t\right) =
\begin{cases}
\left(\frac {\mathbf{Q}_{t-1}^* - \mathbf{Q}_{t}^*} {\tau}, \mathbf{s}_t^*\right)& 1 \leq t \leq p^*,\\
\left(\mathbf{0}, \mathbf{0}\right)& p^* < t \leq T,
\end{cases}
\end{align}
where $(\mathbf{s}_t^*)_{t=1}^{p^*}$, is an optimal solution to \probref{prob:Transmission-Time-Minimization Problem} and $\mathbf{Q}_t^*$ is the corresponding data queue vector in time slot $k$ when applying the optimal solution.
\item   If ${\boldsymbol \mu}_{[T]} \not\in \Lambda_{[T]}$, then solving \probref{prob:Transmission-Time-Minimization Problem} gives $p^* > T$.
\end{enumerate}
\end{theorem}

\begin{IEEEproof}
i) $\forall {\boldsymbol \mu}_{[T]} \in \Lambda_{[T]}$, then the data queue can be cleared with some $p \leq T$, which implies $p^* \leq p \leq T$ holds.
Based on $p^* \leq T$, we prove that~\eqref{eqn:Rate-Achieving Policy for All Achievable Rates} is exactly the rate-achieving policy for ${\boldsymbol \mu}_{[T]}$.
By~\eqref{eqn:Rate-Achieving Policy for All Achievable Rates}, the average rate over $T$ slots is
\begin{align}\label{eqn:Average Rate by Rate-Achieving Policy for All Achievable Rates}
\frac {1} {T} \sum_{t=1}^{T} \frac {\mathbf{Q}_{t-1}^* - \mathbf{Q}_{t}^*} {\tau} = \frac {\mathbf{Q}_{0}} {\tau T} = \frac {\tau T {\boldsymbol \mu}_{[T]}} {\tau T} = {\boldsymbol \mu}_{[T]},
\end{align}
which means the rate is achieved by rate sequence $({(\mathbf{Q}_{t-1}^* - \mathbf{Q}_{t}^*)}/{\tau})_{t=1}^{p^*}$.
Additionally, since the following holds for every $t \in \{1,\ldots,p^*\}$
\begin{align}
\frac {\mathbf{Q}_{t-1}^* - \mathbf{Q}_{t}^*} {\tau} \preceq \mathbf{C} (\mathbf{s}_t^*),
\end{align}
the capacity constraints (see \defref{def:Rate-Achieving Policy}) are satisfied.
Therefore, ${\boldsymbol \mu}_{[T]}$ can be achieved by the policy $\mathcal{P}_T$.

ii) $\forall {\boldsymbol \mu}_{[T]} \not\in \Lambda_{[T]}$, $p > T$ always holds, so does $p^* > T$.
\end{IEEEproof}

\begin{remark} This theorem implies that by solving \probref{prob:Transmission-Time-Minimization Problem} for any given rate, we are able to: (i) directly tell whether the rate is achievable or not by looking at the value of the optimal objective of \probref{prob:Transmission-Time-Minimization Problem}; and (ii) obtain the rate-achieving policy in a closed form based on the solution to \probref{prob:Transmission-Time-Minimization Problem}, if the rate is achievable. Hence, the complexity of finding the rate-achieving policy is the same as that of solving \probref{prob:Transmission-Time-Minimization Problem}.
\end{remark}

\begin{figure}[h]
\centering
\includegraphics [width=0.6\columnwidth]{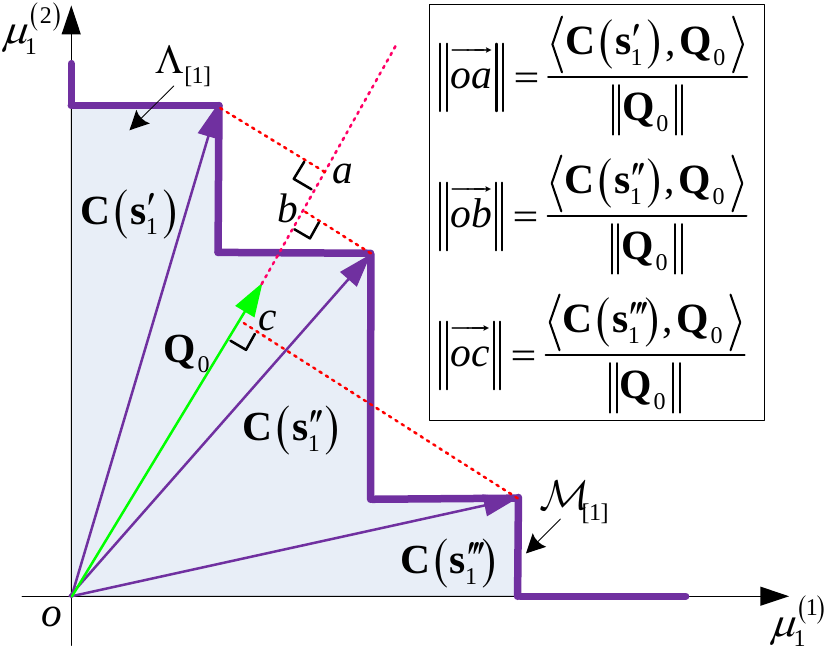}
\caption{The application of the max-weight algorithm in one time slot.
For two transmitter-receiver pairs, we have ${\boldsymbol \mu}_1 = \big[\mu^{(1)}_1,\mu^{(2)}_1\big]^{\mathrm{tr}}$ and the one-slot throughput region $\Lambda_{[1]}$ is a two-dimensional region.
The purple (thick) line segments stand for the weak Pareto frontier $\mathcal{M}_{[1]}$ for $\Lambda_{[1]}$, and the Pareto frontier is $\overline{\mathcal{M}}_{[1]} = \{\mathbf{C}(\mathbf{s}'_1),\mathbf{C}(\mathbf{s}''_1),\mathbf{C}(\mathbf{s}'''_1)\}$.
Without loss of generality and for the simplicity of analysis, we assume the length of a time slot $\tau = 1$.
%
}
\label{fig:Using max-weight algorithm in one time slot}
\end{figure}

\begin{remark}\label{rek:The Reason Why Max-Weight Algorithm Is Not Complete}
The max-weight algorithm\footnote{This algorithm was given in a seminal work in~\cite{TassiulasL1992} and it can achieve all rates in the interior of the infinite-horizon throughput region (not including the boundary rate).} is a commonly used method to find rate-achieving policies over an infinite time horizon.
A natural question is: can we use the max-weight algorithm to derive the rate-achieving policy over a finite horizon of $T$ time slots?
We claim that the max-weight algorithm cannot always give feasible rate-achieving policies for achievable rates over a finite horizon.

A simple and explicit example is given in \figref{fig:Using max-weight algorithm in one time slot}, which illustrates that the max-weight algorithm is not complete in finding rate-achieving policy even for one-slot throughput region:
Assume that we want achieve a rate ${\boldsymbol \mu}_{[1]}$ within one time slot, i.e., $T = 1$.
To achieve this rate, the designed algorithm should find a pair $({\boldsymbol \mu}_1, \mathbf{s}_1)$ such that ${\boldsymbol \mu}_{[1]} = {\boldsymbol \mu}_1 \preceq \mathbf{C}(\mathbf{s}_1)$ holds.
However, the max-weight algorithm cannot return such a rate.
This is because it always sets the transmission rate in a single time slot to be the one having the maximum inner product with the remaining virtual data queue. In the special case of $T = 1$, the remaining virtual data queue is $\mathbf{Q}_0 = {\boldsymbol \mu}_{[1]}$.
From \figref{fig:Using max-weight algorithm in one time slot}, we can see that ${\boldsymbol \mu}_1 = \mathbf{C}(\mathbf{s}'_1)$ will be selected as the transmission rate, since it has the maximum projection $\|\protect\overrightarrow{oa}\|$ on $\mathbf{Q}_0$. However, transmitting at the rate of $\mathbf{C}(\mathbf{s}'_1)$ cannot achieve ${\boldsymbol \mu}_{[1]}$, or more precisely, the required rate of the first transmitter-receiver pair is not achieved. In contrast, setting the transmission rate to $\mathbf{C}(\mathbf{s}''_1)$ is sufficient to achieve ${\boldsymbol \mu}_{[1]}$ (recall that $\mathbf{Q}_0 = {\boldsymbol \mu}_{[1]}$), even though $\mathbf{C}(\mathbf{s}''_1)$ has a smaller projection on $\mathbf{Q}_0$, because $\|\protect\overrightarrow{ob}\| < \|\protect\overrightarrow{oa}\|$.

Therefore, the max-weight algorithm does not always work even for the simplest case of one time slot. The same argument can be extended to examine the transmission policy returned by the max-weight algorithm in the final time slot of a general $T$-slot scenario. Therefore, we conclude that the max-weight algorithm is not complete in finding rate-achieving policies for any finite-horizon throughput region.
\end{remark}


%
%

\subsection{Solving the Transmission-Time-Minimization Problem}\label{sec:Solution Method for Transmission-Time-Minimization Problem}

In \secref{sec:Deriving Rate-Achieving Policy}, the results are based on the solution to \probref{prob:Transmission-Time-Minimization Problem}.
%
%
In this subsection, we discuss how to efficiently solve \probref{prob:Transmission-Time-Minimization Problem}.

To solve~\eqref{eqn:Transmission-Time-Minimization Problem} in \probref{prob:Transmission-Time-Minimization Problem}, intuitively, we could use dynamic programming to search from $\mathbf{Q}_p = \mathbf{0}$ to $\mathbf{Q}_0 = \tau T{\boldsymbol \mu}_{[T]}$ (backward) or employ other uninformed search strategies~\cite{RussellS2009BOOK}.
However, in such searching methods, the number of leaf nodes in the search tree grows exponentially with the depth of the tree and has a large branch factor.
To be more specific, the branching factor is $|\mathcal{S}|$. For example, if we start the search from $\mathbf{Q}_0 = \tau T {\boldsymbol \mu}_{[T]}$, for the first step, we need to calculate all
\begin{align}\label{eqn:Example for The Leaf Nodes in Depth 1}
\mathbf{Q}_1 = \left(\mathbf{Q}_0 - \tau \mathbf{C} \left(\mathbf{s}_1\right)\right)^+,
\end{align}
for all $\mathbf{s}_1 \in \mathcal{S}$.
Thus, the number of leaf nodes is $|\mathcal{S}|$ for the depth $t = 1$.
Similarly, for every possible $\mathbf{Q}_1$ in~\eqref{eqn:Example for The Leaf Nodes in Depth 1}, we have $|\mathcal{S}|$ possible $\mathbf{Q}_2$, and thus the number of leaf nodes for $t = 2$ is $|\mathcal{S}|^2$.
As such, the number of leaf nodes for depth $t = p^*$ (the optimal transmission time) is $|\mathcal{S}|^{p^*}$.
The complexity of such searching methods is $O(|\mathcal{S}|^{p^*})$.

In this subsection, we use the following three steps to significantly improve the computational efficiency in solving  \probref{prob:Transmission-Time-Minimization Problem} and arrive at an lower complexity $O(|B|^{p^*})$, where $B$ is very small compared to $|\mathcal{S}|$.

\textit{Step 1}: Firstly, we reduce the branching factor from $|\mathcal{S}|$ to $|\overline{\mathcal{M}}_{[1]}|$, which is given in \propref{prop:Optimization for the Branching Factor}.
%
%
%
%
%
%
%

\begin{proposition}[Branching Factor Reduction]\label{prop:Optimization for the Branching Factor}
There exists a sequence $(\mathbf{s}_t)_{t=1}^{p^*}$, where $\mathbf{C} \left(\mathbf{s}_t\right) \in \overline{\mathcal{M}}_{[1]},\,t\in\{1,\ldots,p^*\}$, such that $(\mathbf{s}_t)_{t=1}^{p^*}$ is an optimal solution of \probref{prob:Transmission-Time-Minimization Problem}.
\end{proposition}

\begin{IEEEproof}
Let $(\mathbf{s}^*_t)_{t=1}^{p^*}$ be an optimal solution of \probref{prob:Transmission-Time-Minimization Problem}, we have $\mathbf{Q}_{p^*} = \mathbf{0}$, which implies
\begin{align}\label{eqn:Data Cleared Equation}
\tau T {\boldsymbol \mu}_{[T]} \preceq \tau \sum_{t=1}^{p^*} \mathbf{C} (\mathbf{s}_t^*).
\end{align}
Let $(\mathbf{s}_t)_{t=1}^{p^*}$ be the sequence with $\mathbf{C}(\mathbf{s}_t) \in \overline{\mathcal{M}}_{[1]}$, and $\mathbf{C}(\mathbf{s}_t^*) \preceq \mathbf{C}(\mathbf{s}_t)$ ($t\in\{1,\ldots,p^*\}$).
Thus,~\eqref{eqn:Data Cleared Equation} can be rewritten as
\begin{align}
\tau T {\boldsymbol \mu}_{[T]} \preceq \tau \sum_{t=1}^{p^*} \mathbf{C} (\mathbf{s}_t^*) \preceq \tau \sum_{t=1}^{p^*} \mathbf{C} (\mathbf{s}_t),
\end{align}
which implies $\mathbf{Q}_{p^*} = \mathbf{0}$ when applying $(\mathbf{s}_t)_{t=1}^{p^*}$.
Therefore, $(\mathbf{s}_t)_{t=1}^{p^*}$ is an optimal solution of \probref{prob:Transmission-Time-Minimization Problem}.
\end{IEEEproof}

\begin{remark}\label{rek:Reduction in the Branching Factor}
\propref{prop:Optimization for the Branching Factor} tells that we only need to consider the transmit powers corresponding to the rate on the Pareto frontier of the one-slot throughput region, instead of all possible transmit powers. Hence, the transmit-power-vector set $\mathcal{S}$ in \probref{prob:Transmission-Time-Minimization Problem} can be substituted by $\overline{\mathcal{S}}$, called the refined transmit-power-vector set, such that $\mathbf{C}(\mathbf{s}_t) \in \overline{\mathcal{M}}_{[1]}$ holds for all $\mathbf{s}_t \in \overline{\mathcal{S}}$.
Therefore, the branching factor is $|\overline{\mathcal{S}}| = |\overline{\mathcal{M}}_{[1]}|$.
\end{remark}

\textit{Step 2}: More importantly, A* search is employed to further improve the searching efficiency while maintaining the optimality (see~\cite{RussellS2009BOOK}) for \probref{prob:Transmission-Time-Minimization Problem}.
A brief description is given here on the application of A* search in solving \probref{prob:Transmission-Time-Minimization Problem}, while we refer the readers to~\cite{RussellS2009BOOK} for a complete description of the A* search algorithm.

For A* search (or any searching algorithm in general), `node' is a fundamental concept.
In our case, a node is given by $\left(\mathbf{Q}_t, (\mathbf{s}_i)_{i=1}^{t}\right)$, which depends on $\mathbf{Q}_t$ the state, and $(\mathbf{s}_i)_{i=1}^{t}$ the path to achieve this state from the initial node $\left(\mathbf{Q}_0, \emptyset\right)$.
A* search requires five components to be implemented:
\begin{itemize}
\item   \textbf{Initial node.} The node starting the search, which is $\left(\mathbf{Q}_0, \emptyset\right)$.
\item   \textbf{Action space.} The set of actions that move from a node to all possible child nodes.
    In our case, the action space is $\overline{\mathcal{S}}$ according to \rekref{rek:Reduction in the Branching Factor}.
\item   \textbf{Goal.} The condition for stopping the search.
    In our case, the goal is $\mathbf{Q}_p = \mathbf{0}$ or simply denoted as $\mathbf{0}$.
\item   \textbf{Step cost.} The step cost is the cost for each searching step.
    In \probref{prob:Transmission-Time-Minimization Problem}, the step cost is $c_t = 1,t \in \{1,\ldots,p\}$.
    %
    %
\item   \textbf{Evaluation function.} It records the path cost (the summation of all previous step costs) from the past and estimates the path cost in the future.
To be more specific, for a given node $\left(\mathbf{Q}_t, (\mathbf{s}_i)_{i=1}^{t}\right)$, the evaluation function is
\begin{align}\label{eqn:Knowledge-Plus-Heuristic Cost Function}
F\left(\mathbf{Q}_t, (\mathbf{s}_i)_{i=1}^{t}\right) = G\left((\mathbf{s}_i)_{i=1}^{t}\right) + H(\mathbf{Q}_t),
\end{align}
where $G\left((\mathbf{s}_i)_{i=1}^{t}\right)$ returns the path cost from the initial node to node $\left(\mathbf{Q}_t, (\mathbf{s}_i)_{i=1}^{t}\right)$. A heuristic function $H(\mathbf{Q}_t)$ estimates the path cost from $\left(\mathbf{Q}_t, (\mathbf{s}_i)_{i=1}^{t}\right)$ to the goal $\mathbf{0}$.
A* search always expands the node with smallest $F$.
\end{itemize}

It should be noted that the core of A* search is to construct an admissible heuristic function\footnote{That is, $H(\mathbf{Q}_t) \leq H^*(\mathbf{Q}_t)$ holds for every $\mathbf{Q}_t$, where $H^*(\mathbf{Q}_t)$ is the actual cost from $\mathbf{Q}_t$ to the goal $\mathbf{0}$.}, since other parts of A* can be determined by the definition of the problem.
In this work, we propose the interference-free based heuristic function:
\begin{align}\label{eqn:Interference-Free Based Heuristic Function}
H^I\left(\mathbf{Q}_t\right) = \max_{n\in \mathcal{N}} \frac {Q_t^{(n)}} {\tau \log_2(1 + \gamma'_n(s_{\max}^{(n)}))},
\end{align}
where $t\in\{1,\ldots,p\}$, $s_{\max}^{(n)} := \max \mathcal{S}^{(n)}$, and
\begin{align}\label{eqn:Interference Free Gamma}
\gamma'_n(s_{\max}^{(n)}) = \frac {h_{nn} s_{\max}^{(n)}} {W_n}, \, n\in\mathcal{N}.
\end{align}
This heuristic function is interference-free based, since compared to~\eqref{eqn:SINR},~\eqref{eqn:Interference Free Gamma} does not consider the interference from other transmitters.
The following proposition states that $H^I(\mathbf{Q}_t)$ is admissible, which means A* search can be employed.

\begin{proposition}[Admissibility of Interference-Free Based Heuristic Function for \probref{prob:Transmission-Time-Minimization Problem}]\label{prop:Admissibility of Interference-Free Based Heuristic Function for Problem 1}
Let the actual cost to reach the goal $\mathbf{Q}_p = \mathbf{0}$ be $H^*\left(\mathbf{Q}_t\right) = p - t$, where $t \in \{1,\ldots,p\}$.
Then $H^I\left(\mathbf{Q}_t\right) \leq H^*\left(\mathbf{Q}_t\right)$ holds for every $\mathbf{Q}_t$.
\end{proposition}

\begin{IEEEproof}
$\forall \mathbf{Q}_t$, let $\overline{\mathbf{s}}_k = \big[\overline{s}_k^{(1)},\ldots,\overline{s}_k^{(n)}\big],\, k \in \{t+1,\ldots,p\}$ be any possible action (transmit-power vector) after $\mathbf{Q}_{k-1}$.
$\forall n \in \mathcal{N}$, we have
\begin{align}\label{eqn:Actual Cost to Reach the Goal Part 1}
H^*\left(\mathbf{Q}_t\right) \!=\! p - t \!=\! \sum_{k = t + 1}^{p} \!1 \!\geq\! \sum_{k = t + 1}^{p} \frac {Q_{k-1}^{(n)} - Q_{k}^{(n)}} {\tau \log_2 \left(1 + \gamma_n (\overline{\mathbf{s}}_k)\right)}.
\end{align}
Additionally, since $\overline{s}_k^{(n)} \leq s_{\max}^{(n)}$, the following holds
\begin{align}
\gamma_n (\overline{\mathbf{s}}_k) = \frac {h_{nn} \overline{s}_k^{(n)}} {W_n + \sum_{m\neq n} h_{mn} \overline{s}_k^{(m)}} \leq \frac {h_{nn} s_{\max}^{(n)}} {W_n} = \gamma'_n(s_{\max}^{(n)}).
\end{align}
Thus,~\eqref{eqn:Actual Cost to Reach the Goal Part 1} can be further bounded from below as
%
\begin{align}\label{eqn:Actual Cost Part 1 Zoom Final}
\begin{split}
H^*\left(\mathbf{Q}_t\right) \!&\geq \max_{n\in \mathcal{N}} \sum_{k = t + 1}^{p} \frac {Q_{k-1}^{(n)} - Q_{k}^{(n)}} {\tau \log_2 \left(1 + \gamma'_n (s_{\max}^{(n)})\right)}\\
&= \max_{n\in \mathcal{N}} \left\{\frac {Q_t^{(n)} - Q_{p}^{(n)}} {\tau\log_2(1 + \gamma'_n(s_{\max}^{(n)}))}\right\} \!=\! H^I\left(\mathbf{Q}_t\right).
\end{split}
\end{align}
Therefore, $H^I\left(\mathbf{Q}_t\right) \leq H^*\left(\mathbf{Q}_t\right)$ holds.
\end{IEEEproof}


\textit{Step 3}: Last but not least, we propose a pruning strategy to further improve the searching efficiency of A* search:
After selecting a node to expand, labelled by $(\mathbf{Q}_{t_1},(\mathbf{s}_i)_{i=1}^{t_1})$, we delete those nodes with $t \geq t_1$ but $(\mathbf{C}(\mathbf{s}_i))_{i=1}^{t} \preceq (\mathbf{C}(\mathbf{s}_i))_{i=1}^{t_1}$ in the fringe (or called open set, more details can be found in~\cite{RussellS2009BOOK}), since those nodes' subtrees are suboptimal or can be replaced with the new node $(\mathbf{Q}_{t_1},(\mathbf{s}_i)_{i=1}^{t_1})$.

To sum up, the method for solving \probref{prob:Transmission-Time-Minimization Problem} is given in~\algref{alg:Solving Problem 1 with A* Search}, in which our pruning strategy is implicitly included in the A* search algorithm.

\begin{algorithm}
\begin{footnotesize}
\caption{Solving \probref{prob:Transmission-Time-Minimization Problem} with A* Search}\label{alg:Solving Problem 1 with A* Search}
\begin{algorithmic}[1]
\REQUIRE
    $T$: number of time slots; $N$: the number of transmitter-receiver pairs; ${\boldsymbol \mu}_{[T]}$: the given rate to be achieved; $\overline{\mathcal{S}}$: refined transmit-power-vector set.
\ENSURE
    $(\mathbf{s}_t^*)_{t=1}^{p^*}$: the optimal solution of \probref{prob:Transmission-Time-Minimization Problem};\\
    $p^*$: the optimal objective of \probref{prob:Transmission-Time-Minimization Problem}.
\STATE  $\mathbf{Q}_0 = \tau T {\boldsymbol \mu}_{[T]}$;
\STATE  $\left[(\mathbf{s}_t^*)_{t=1}^{p^*}, p^*\right] = \mathrm{A}^*\left(\left(\mathbf{Q}_0, \emptyset\right), \overline{\mathcal{S}}, \mathbf{0}, c_t, F(\cdot)\right)$;
\RETURN $(\mathbf{s}_t^*)_{t=1}^{p^*}$ and $p^*$.
\end{algorithmic}
\end{footnotesize}
\end{algorithm}

We use the concept of effective branching factor\footnote{It is a popular metric for characterizing the quality of searching method, e.g., see Section 3.6.1 in~\cite{RussellS2009BOOK}.} (EBF) to measure the searching efficiency of the proposed solution to \probref{prob:Transmission-Time-Minimization Problem}.
For a fixed $p^*$, the relationship between $U$ (the total number of expanded nodes) and the EBF is
\begin{align}\label{eqn:U}
U = \sum_{t=1}^{p^*} B^t.
\end{align}
where $B$ is the EBF. We can see that $B$ polynomially increases with $U$, which means the smaller the EBF is, the better our algorithm performs.
In \secref{sec:Simulation Examples}, we will present numerical results on EBF to measure the searching efficiency.

\section{Numerical Results}\label{sec:Simulation Examples}

In this section, we present numerical results to corroborate our analytical results.
First, we give two illustrative examples with different channel conditions: an example of a given rate falling in the throughput region (i.e., achievable rate) and an example of a given rate falling out of the throughput region. Consider a network with $N = 3$ transmitter-receiver pairs within $T = 5$ time slots.
The transmit-power sets of these $3$ transmitter-receiver pairs are $\mathcal{S}^{(1)} = \mathcal{S}^{(2)} = \mathcal{S}^{(3)} = \{0, 2\}$, which actually represent an on-off transmission scheme.
The noise powers are $W_1 = W_2 = W_3 = 0.1$, and the length of a time slot $\tau$ is normalized to $1$.
Under the following two different channel conditions, we want to achieve the rate ${\boldsymbol \mu}_{[5]} = [1, 1, 1]^{\mathrm{tr}}$:
\begin{itemize}
\item   Consider channel power gains $h_{11} = 0.5$, $h_{22} = 0.6$, $h_{33} = 0.7$, and $h_{12} = h_{21} = h_{13} = h_{31} = h_{23} = h_{32} = 0.2$.
    By using \thmref{thm:Rate-Achieving Policy for All Achievable Rates} and solving \probref{prob:Transmission-Time-Minimization Problem} with the proposed A* search algorithm, the rate-achieving policy is $\mathcal{P}_5 = ({\boldsymbol \mu}_t,\mathbf{s}_t)_{t=1}^5$, where ${\boldsymbol \mu}_1 = [3.4594,0,0]^{\mathrm{tr}}$, ${\boldsymbol \mu}_2 = {\boldsymbol \mu}_3 = [0,1.7655,1.9260]^{\mathrm{tr}}$, ${\boldsymbol \mu}_4 = [1.0780,1.2224,1.1480]^{\mathrm{tr}}$, ${\boldsymbol \mu}_5 = [0.4626,10.2465,0]^{\mathrm{tr}}$, and $\mathbf{s}_1 = [2,0,0]^{\mathrm{tr}}$, $\mathbf{s}_2 = \mathbf{s}_3 = [0,2,2]^{\mathrm{tr}}$, $\mathbf{s}_4 = [2,2,2]^{\mathrm{tr}}$, $\mathbf{s}_5 = [2,2,0]^{\mathrm{tr}}$.
\item   Consider channel power gains $h_{11} = h_{22} = h_{33} = 0.2$, and $h_{12} = h_{21} = h_{13} = h_{31} = h_{23} = h_{32} = 0.5$.
    Solving \probref{prob:Transmission-Time-Minimization Problem} gives $p^* =8 > 5$. Hence,
    \thmref{thm:Rate-Achieving Policy for All Achievable Rates} tells that the rate $[1, 1, 1]^{\mathrm{tr}}$ is not achievable in $T = 5$ time slots.
\end{itemize}

Next, we conduct Monte Carlo simulations to examine the computational efficiency.
All the system parameters including the given rate to be achieved ${\boldsymbol \mu}_{[5]} = [1, 1, 1]^{\mathrm{tr}}$ remain the same instead of the channel power gains. Here, we consider many possible realizations of the fading channels. Specifically, we use Nakagami-$m$ fading with $m \in \{1,\ldots,5\}$ to generate $10000$ realizations of the channel for each communication and interference link (hence we have $10000$ different scenarios). The average EBF (effective branching factor) is given in \tabref{tab:Average EBF}. Assuming $p^* = T = 5$, then the average total number of nodes (except for the starting node) of the original tree (computed using~\eqref{eqn:U}) without applying any of the three steps in \secref{sec:Main Results} is $\sum_{t=1}^5 8^t = 37449$. But using our A* search with pruning, e.g., for $m = 3$, the average number of expanded nodes is only $\sum_{t=1}^5 3.6116^t \approx 849$.
This shows a significant improvement in the computational efficiency.

\begin{table}
\caption{Average EBF under different Nakagami-$m$ fading\label{tab:Average EBF}}
\centering
\begin{small}
\begin{tabular}{lccccc}
\toprule
~& $m = 1$ & $m = 2$ & $m = 3$ & $m = 4$ & $m = 5$\\
\midrule
EBF & $3.5557$ & $3.5757$ & $3.6116$ & $3.6334$ & $3.6502$\\
\bottomrule
\end{tabular}
\end{small}
\end{table}

\section{Conclusion}

For the first time, this work studied the throughput region of a wireless multi-user interference channel over a finite time horizon. We provided a computationally efficient algorithm that determines whether a rate is achievable in a given finite number of time slots, and if so this algorithm provides the rate-achieving policy (a sequence of rate-power pairs) to achieve that rate. We started by formulating an equivalent transmission-time-minimization problem whose optimal solution provides a closed-form expression for the rate-achieving policy. In order to efficiently solve the transmission-time-minimization problem, we applied three steps:
i) branch factor reduction;
ii) A* search algorithm with a carefully chosen admissible heuristic function; and
iii) pruning strategy.
Simulation results demonstrated the efficiency of the proposed method in improving the computational efficiency.

%

\bibliographystyle{IEEEtran}

\bibliography{FHTR}

\begin{thebibliography}{10}
\providecommand{\url}[1]{#1}
\csname url@samestyle\endcsname
\providecommand{\newblock}{\relax}
\providecommand{\bibinfo}[2]{#2}
\providecommand{\BIBentrySTDinterwordspacing}{\spaceskip=0pt\relax}
\providecommand{\BIBentryALTinterwordstretchfactor}{4}
\providecommand{\BIBentryALTinterwordspacing}{\spaceskip=\fontdimen2\font plus
\BIBentryALTinterwordstretchfactor\fontdimen3\font minus
  \fontdimen4\font\relax}
\providecommand{\BIBforeignlanguage}[2]{{%
\expandafter\ifx\csname l@#1\endcsname\relax
\typeout{** WARNING: IEEEtran.bst: No hyphenation pattern has been}%
\typeout{** loaded for the language `#1'. Using the pattern for}%
\typeout{** the default language instead.}%
\else
\language=\csname l@#1\endcsname
\fi
#2}}
\providecommand{\BIBdecl}{\relax}
\BIBdecl

\bibitem{NeelyM2005JSAC}
M.~Neely, E.~Modiano, and C.~Rohrs, ``Dynamic power allocation and routing for
  time-varying wireless networks,'' \emph{IEEE J. Sel. Areas Commun.}, vol.~23,
  no.~1, pp. 89--103, Jan. 2005.

\bibitem{GeorgiadisL2006BOOK}
L.~Georgiadis, M.~J. Neely, and L.~Tassiulas, \emph{Resource Allocation and
  Cross-Layer Control in Wireless Networks}.\hskip 1em plus 0.5em minus
  0.4em\relax Found. Trends Netw., 2006.

\bibitem{VaeziM2016ICC}
M.~Vaezi and H.~V. Poor, ``Simplified han-kobayashi region for one-sided and
  mixed gaussian interference channels,'' in \emph{Proc. IEEE Int. Conf. on
  Commun. (ICC)}, Kuala Lumpur, Malaysia, May 2016.

\bibitem{TassiulasL1992}
L.~Tassiulas and A.~Ephremides, ``Stability properties of constrained queueing
  systems and scheduling policies for maximum throughput in multihop radio
  networks,'' \emph{IEEE Trans. Autom. Control}, vol.~37, no.~12, pp.
  1936--1948, Dec. 1992.

\bibitem{TassiulasL1991Thesis}
L.~Tassiulas, ``Dynamic link activation scheduling in multihop radio networks
  with fixed or changing connectivity,'' Ph.D. dissertation, University of
  Maryland, College Park, MD, USA, 1991.

\bibitem{LinX2006}
X.~Lin, N.~Shroff, and R.~Srikant, ``A tutorial on cross-layer optimization in
  wireless networks,'' \emph{IEEE J. Sel. Areas Commun.}, vol.~24, no.~8, pp.
  1452--1463, Aug. 2006.

\bibitem{NeelyM2010}
M.~J. Neely, ``Stochastic network optimization with application to
  communication and queueing systems,'' \emph{Synthesis Lectures on Commun.
  Netw.}, vol.~3, no.~1, pp. 1--211, 2010.

\bibitem{LeL2012}
L.~B. Le, E.~Modiano, and N.~Shroff, ``Optimal control of wireless networks
  with finite buffers,'' \emph{IEEE/ACM Trans. Netw.}, vol.~20, no.~4, pp.
  1316--1329, Aug. 2012.

\bibitem{XueD2013VT}
D.~Xue and E.~Ekici, ``Power optimal control in multihop wireless networks with
  finite buffers,'' \emph{IEEE Trans. Veh. Technol.}, vol.~62, no.~3, pp.
  1329--1339, Mar. 2013.

\bibitem{XueD2015}
D.~Xue, R.~Murawski, and E.~Ekici, ``Capacity achieving distributed scheduling
  with finite buffers,'' \emph{IEEE/ACM Trans. Netw.}, vol.~23, no.~2, pp.
  519--532, Apr. 2015.

\bibitem{RussellS2009BOOK}
S.~Russell, \emph{Artificial intelligence: A modern approach}, 3rd~ed.\hskip
  1em plus 0.5em minus 0.4em\relax Prentice Hall, 2009.

\bibitem{TanC2013}
C.~W. Tan, M.~Chiang, and R.~Srikant, ``Fast algorithms and performance bounds
  for sum rate maximization in wireless networks,'' \emph{IEEE/ACM Trans.
  Netw.}, vol.~21, no.~3, pp. 706--719, June 2013.

\bibitem{MotahariA2009}
A.~Motahari and A.~Khandani, ``Capacity bounds for the gaussian interference
  channel,'' \emph{IEEE Trans. Inf. Theory}, vol.~55, no.~2, pp. 620--643, Feb.
  2009.

\end{thebibliography}
\end{document}